\documentclass[conference]{IEEEtran}

\usepackage{myStyleIEEE}
\usepackage{ amssymb }
\usepackage{mathtools}

\IEEEoverridecommandlockouts

\begin{document}

\title{A Linear Formulation for Power System State Estimation including RTU and PMU Measurements  \vspace*{-2mm}}

\renewcommand{\theenumi}{\alph{enumi}}
\renewcommand{\arraystretch}{1.3}
\author{
\IEEEauthorblockN{Aleksandar Jovicic\IEEEauthorrefmark{1}, Marko Jereminov\IEEEauthorrefmark{2}, Larry Pileggi\IEEEauthorrefmark{2}, Gabriela Hug\IEEEauthorrefmark{1}}%
\IEEEauthorblockA{\IEEEauthorrefmark{1} EEH - Power Systems Laboratory, ETH Zurich, Zurich, Switzerland } %
\IEEEauthorblockA{\IEEEauthorrefmark{2} Department of Electrical and Computer Engineering, Carnegie Mellon University, Pittsburgh, PA\\
Emails: \{jovicic, hug\}@eeh.ee.ethz.ch, \{mjeremin, pileggi\}@andrew.cmu.edu \vspace*{-0.525cm}}

}

\maketitle
\IEEEpeerreviewmaketitle

            
\begin{abstract}
In this paper, a novel linear formulation for power system state estimation that simultaneously treats conventional and synchrophasor measurements is proposed. A linear circuit model for conventional measurements is introduced to enable a fully linear equivalent circuit representation of the power system. The estimated system state is then obtained by formulating the optimization problem to minimize the measurement errors and solving the resulting linear set of optimality conditions. To evaluate the accuracy of the proposed method, simulations are performed on several test cases of various sizes and the results are presented and discussed. \\
\end{abstract}

\begin{IEEEkeywords}
State estimation, equivalent circuit formulation, phasor measurement units, conventional measurements, linear estimation
\end{IEEEkeywords}

\section{Introduction}
The present day power systems are undergoing significant changes due to increased penetration of distributed sources, diversified nature of loads and new market strategies. In order to operate the grid reliably and address all these novelties, a necessary prerequisite is to have an accurate insight into the actual operating state of the grid at each point in time. State Estimation (SE) is a mathematical tool that provides an estimate of the most probable state of the system based on the raw measurement data collected from the monitoring devices installed throughout the network. The system state is traditionally represented as a set of voltage magnitudes and angles for each bus in the system \cite{Abur}. 

The formulation of any SE algorithm heavily depends on the type of available measurements. For a long time, Remote Terminal Units (RTU) were the main source of measurement data, providing measurements of bus voltage magnitudes and active and reactive power flows and injections. Therefore, all SE methods that have been developed until recently were based on these data \cite{Abur}. The Weighted Least Square (WLS) method, formulated in \cite{Schweppe}, is one of the most extensively used approaches in the SE area. The measurement functions used in this approach are derived from the power flow equations, resulting in an inherently nonlinear problem. 

Phasor Measurement Units (PMU) are an emerging technology that has substantially improved the capability to monitor the state of the system. These devices can provide highly accurate measurements of current and voltage phasors, which are time-synchronized and communicated to the grid operator at a much higher time resolution compared to the RTU measurements. Having a system that is fully observable solely by PMUs is an ideal scenario, since the state of the system would be determined with a very high precision in such case. Furthermore, the SE problem would become linear, under the assumption that the state vector and measurement functions in the conventional WLS are formulated in terms of voltages and currents in rectangular coordinates \cite{Phadke1985}. However, this scenario is unrealistic at the moment, mainly due to the significant costs of PMU devices and the penetration of legacy measuring equipment. Therefore, algorithms that can combine both conventional and synchrophasor measurements are required.

Many algorithms that address the problem of simultaneous treatment of PMU and RTU measurements can be found in literature and are generally categorized into two different groups. Hybrid estimators that treat different types of measurements at separate stages are proposed in \cite{Ming2006,Baltensperger2010,Nuqui}. These approaches assume full system observability by RTUs and utilize nonlinear measurement functions based on power flow equations. The drawbacks of multi-stage estimators are eliminated to a certain extent by the introduction of single-stage estimators which treat PMU and RTU measurements simultaneously \cite{Bi2008,Chakrabarti2010,Valverde2011}. However, all of these methods rely on a WLS problem formulation which has to be solved iteratively due to its nonlinearity.        

In this paper, we propose an approach that is derived from a recently introduced Equivalent Circuit Formulation (ECF) for the power flow problem \cite{CMU2,CMU3,CMU4,Pandey}. The idea of this power flow approach is that a power system can be more naturally described in terms of voltage and current state variables in rectangular coordinates and further represented by an equivalent split-circuit. This allows for numerous circuit simulation tools that have been developed in the circuit community to be applied to a range of power system problems \cite{CMU4,CMU6}. Furthermore, any arbitrary power system element can be incorporated within the ECF framework based on the relation between voltage and current at its terminals. 

As a consequence, a novel approach for static state estimation based on current, voltage and admittance state variables was proposed in \cite{Jovicic}. Circuit models for synchrophasor and conventional measurements were developed and incorporated into the ECF framework along with the models of different power system elements that were previously derived in \cite{Pandey}. An optimization problem was formulated in order to estimate the state of the system in rectangular coordinates. With the RTU circuit being the only nonlinear element, the proposed method features a significantly lower level of nonlinearity compared to the conventional WLS algorithm. Additionally, PMU and RTU measurements are treated simultaneously. However, due to the nonlinearity of the proposed formulation introduced by modelling the RTU measurements, a nonlinear optimization solver is required.

In this paper, a novel linear formulation for the state estimation problem is derived from the method presented in \cite{Jovicic}. This is achieved by introducing a linear circuit model for conventional measurements rendering the governing equations of the SE problem defined in terms of current and voltage state variables in rectangular coordinates linear. Subsequently, an optimization problem is formulated in order to estimate the states in rectangular coordinates. Contrary to \cite{Jovicic}, measurement uncertainties are accounted for by assigning the appropriate weight coefficients to each term in the objective function as opposed to explicitly assigning boundaries. This removes the explicit bounds, but on the other hand, also removes some nonlinearity within the optimality conditions. The first-order optimality conditions for this problem are derived and shown to be linear. Ultimately, the state of the system is estimated by solving a system of linear equations, thus avoiding any iterative procedures and drastically improving the computational time and scalability of the proposed algorithm.

Section \ref{sec: 2} gives an overview of the equivalent circuit formulation of the power flow problem, along with a brief discussion about the weighted approach for the treatment of measurement uncertainties. In Sect. \ref{sec: 3}, the linear modelling framework for different measurement sets is described and the SE optimization problem is formulated. Section \ref{sec: 4} showcases the simulation results, while the main conclusions are presented in Sect. \ref{sec: 5}.    

\section{Technical Background} \label{sec: 2}
\subsection{Equivalent Circuit Formulation}
An equivalent split-circuit representation of the power flow problem in terms of current and voltage state variables was introduced recently in \cite{CMU2,CMU3,CMU4}. In this formulation, each element of the power system is initially described by the relations between currents and voltages at its terminals and is subsequently translated into a suitable circuit model. The obtained models are then interconnected to construct the equivalent circuit of the entire system. Then, some of the commonly used methods for formulation of circuit equations, such as Modified Nodal Analysis or Tree-Link Analysis \cite{Pileggi}, are used to derive the circuit equations that are solved to obtain the power flow solution in rectangular coordinates.

Many power system elements are nonlinear and mathematically represented by nonanalytic complex equations. Therefore, common linearization techniques, such as the first order Taylor approximation, cannot be applied since the presence of the complex conjugate operator prevents the use of derivatives. In order to solve this issue, the equations that define voltage-current relations of each power system element are split into two sets of equations, where the first set represents the real state variables, and the second represents their imaginary part \cite{CMU2}. This corresponds to splitting the power system equivalent circuit into two coupled circuits, where the first circuit contains only real voltages and currents, while the second comprises their imaginary parts. The coupling of the two split circuits is implemented via controlled sources. In this way, the circuit equations can be linearized and iteratively solved, e.g. by applying the Newton-Raphson method. For more details, the reader is referred to \cite{CMU2,CMU3,CMU4,Pandey}. 

\subsection{Weighted Approach for Measurement Uncertainties} \label{subsec: Intervals}
All available measurement data contain some degree of uncertainty, regardless of the type of measured physical quantity or particular measurement device that is used. The measurement uncertainty is usually quantified in terms of its standard deviation $\sigma$. One of the essential tasks of any SE algorithm is to account for the difference in accuracy of various measurements. The way this is done in conventional WLS method is to assign a weight coefficient to each particular measurement equal to the reciprocal value of the respective measurement variance $\sigma^2$. By doing so, higher weight factors are obtained for the measurements with smaller standard deviations. Therefore, more accurate measurements affect the final estimate of the system state more than those that have larger standard deviations.

For the set of measurements that are directly incorporated into the SE formulation, their respective weights are trivially calculated as reciprocal values of their variances. However, depending on the framework that is used for state estimation, some measured quantities cannot be directly used. Instead, some form of transformation of the original data is needed, which brings up the question of error propagation \cite{Navidi}. In order to calculate the variance of the transformed measurement data, the following equations will be used:
\begin{gather}
    f = A + B \implies \sigma_{f}^2 = \sigma_{A}^2 + \sigma_{B}^2 \label{eq: var_sum}\\ 
    f = AB \,\, \text{or} \,\, f = \frac{A}{B} \implies \sigma_{f}^2 = f^2\left[\left(\frac{\sigma_{A}}{A}\right)^2+\left(\frac{\sigma_{B}}{B}\right)^2\right] \label{eq: var_product} 
\end{gather}
where $A$ and $B$ are the original measurements that are transformed to the pseudo-measurement $f$, while $\sigma_A$, $\sigma_B$ and $\sigma_f$ are standard deviations of measurements $A$ and $B$ and pseudo-measurement $f$, respectively. Measurement errors are assumed to be normally distributed and mutually independent. 

Different from the approach in \cite{Jovicic}, which assumes that the true value is within the measurement bounds, this approach may also incorporate bad measurements. Additionally, weight factors preserve the information about the distribution of measurement errors, while the interval approach assumes uniform error distribution.      

\section{Proposed Linear State Estimator} \label{sec: 3}
Two major steps have to be undertaken in order to achieve a fully linear state estimator based on the equivalent circuit approach. First, all measurements have to be represented in terms of voltage and current state variables, with the aim of developing a fully linear equivalent circuit of the system. Subsequently, the optimization problem with linear first-order optimality conditions is defined in order to estimate the most likely state of the system.
    
\subsection{Linear Circuit Modelling} \label{subsec: 3a}
As it was shown in \cite{CMU2,CMU3,CMU4,Pandey}, any physics based model can be mapped to an equivalent circuit in terms of voltage and current state variables. This implies that any available set of measurements can be translated into the equivalent circuit framework, since different measurement devices installed in the grid essentially measure voltages and currents.

\begin{figure}[h!]
    \vspace{-0.175cm}
    \centering
    \begin{subfigure}[h]{0.135\textwidth}
        \centering  
        \includegraphics[width=\textwidth]{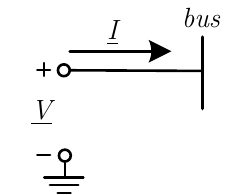}
        \caption{\hspace*{0.25em}}  
        \label{fig:PMU_Inj_a}
    \end{subfigure}
    \hfill
    \begin{subfigure}[h]{0.165\textwidth}  
        \centering 
        \includegraphics[width=\textwidth]{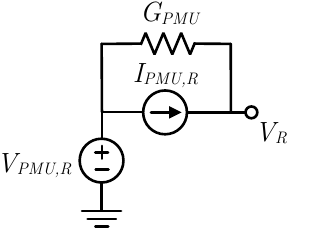}
        \caption{\hspace*{0.25em}}  
        \label{fig:PMU_Inj_b}
    \end{subfigure}
    \hfill
        \begin{subfigure}[h]{0.165\textwidth}  
        \centering 
        \includegraphics[width=\textwidth]{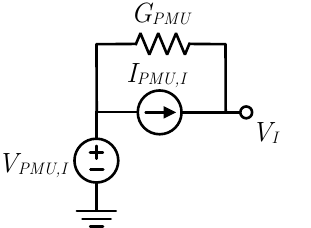}
        \caption{\hspace*{0.25em}}  
        \label{fig:PMU_Inj_c}
    \end{subfigure}
    \caption{PMU injection model: (a) measurement data; (b) real circuit; (c) imaginary circuit.}
    \label{fig:PMU_inj}

\end{figure}

PMU devices measure phasors of voltages and currents. Since the ECF framework is based on voltage and current state variables in rectangular coordinates, the PMU measurements can be trivially mapped to the equivalent circuit as independent voltage and current sources. A circuit model for synchrophasor measurements was already derived in \cite{Jovicic}. The corresponding model is shown in Fig.~\ref{fig:PMU_inj}, and it consists of a voltage source in series with a parallel connection of a current source and a conductance. The sources represent the measured values. 

\begin{figure}[t!] 
\vspace{-1em}
\centering

\begin{subfigure}[h]{.45\linewidth}
\includegraphics[width=\linewidth]{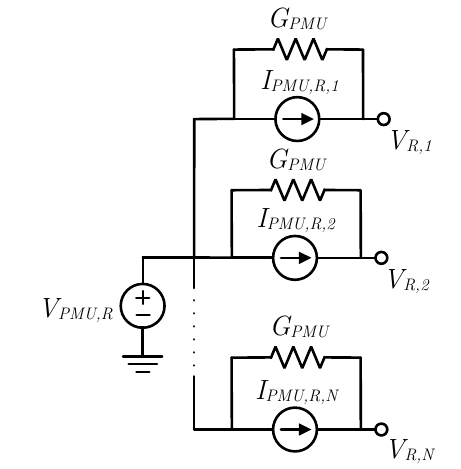}
\caption{\label{fig:PMU_flow_b}}
\end{subfigure}
\begin{subfigure}[h]{.45\linewidth}
\includegraphics[width=\linewidth]{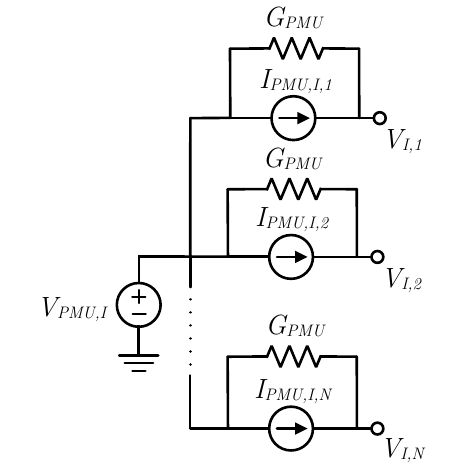}
\caption{\label{fig:PMU_flow_c}}
\end{subfigure}
\caption{PMU line flow model: (a) real circuit; (b) imaginary circuit.}
\label{fig:PMU_flow}
\vspace{-1em}
\end{figure}
\noindent By adding the conductance $G_{PMU}$, the influence of both voltage and current sources is preserved. This would not be the case if these two sources were connected in series without a counductance, because in such case a voltage source would not have any impact. Additionally, the conductances in the PMU sub-circuit also serve as a slack element, because there is a current flowing through these elements only if the voltages and currents representing the inaccurate measurements fail to satisfy the circuit constraints enforced by the rest of the equivalent circuit for the system. This essentially means that the currents flowing through the PMU conductances will be zero in case all measurements in the system are perfectly accurate, which will be leveraged in the optimization problem described in section \ref{subsec: 3b}. A similar modelling approach can be used in case line flow measurements are available instead of the injection current \cite{Jovicic}. Then, each line flow is represented by a parallel connection of a current source and a conductance, as shown in Fig.~\ref{fig:PMU_flow}. In case a line flow measurement is missing in any of the lines, the respective transmission line terminal should be directly connected to the voltage source. Finally, an important observation is that the proposed models for synchrophasor measurements are linear. 

The values of voltage and current sources in PMU sub-circuits are considered as unknown variables. Different from \cite{Jovicic} where measurement errors were treated as intervals, the proposed linear formulation for SE utilizes a weighted approach to account for the uncertainties. A weight factor is assigned to each voltage and current source modelling a particular PMU measurement, and is equal to the reciprocal value of the variance of the respective measurement. This will be utilized in the optimization problem that is elaborated in section \ref{subsec: 3b}.  

Conventional measurements commonly comprise bus voltage magnitudes, as well as active and reactive power flows and injections. However, each RTU device essentially measures voltage and current magnitude and the phase angle between them \cite{Caro}. These quantities are further used to calculate power measurements, which consequently feature higher standard deviations than the originally measured data. Therefore, for the sake of improved accuracy, it is assumed that the set of RTU measurements consists of voltage and current magnitudes, and the phase angle between them. These types of data are not directly compliant with the ECF framework, hence they have to be converted to a form that provides information about relations between voltages and currents in rectangular coordinates. 

Active ($P$) and reactive ($Q$) power can be expressed in terms of the measured bus voltage magnitude ($V$), current magnitude ($I$) and the angle between them ($\phi$) by the following relations:
\begin{align}
    P &= VI\,\text{cos}\,\phi \label{eq: P_RTU}\\
    Q &= VI\,\text{sin}\,\phi \label{eq: Q_RTU}
\end{align}
where the reference directions of voltage and current correspond to load conditions. For the same reference directions, $P$ and $Q$ which are drawn at a bus can be expressed in terms of voltage and current in rectangular coordinates as follows:
\begin{align}
    P &= V_{R}I_{R} + V_{I}I_{I} \label{eq: activeP}\\
    Q &= -V_{R}I_{I} + V_{I}I_{R} \label{eq: reacQ}
\end{align}
where $V_R$ and $V_I$ are real and imaginary voltage, and $I_R$ and $I_I$ are real and imaginary current.  
Finally, equations for real and imaginary current as a function of real and imaginary voltage and RTU measurements are derived as:
\begin{align}
    I_{R} &= \frac{I}{V}\text{cos}\left(\phi\right)V_{R} + \frac{I}{V}\text{sin}\left(\phi\right)V_{I} = I_{GR} + I_{BR} \label{eq: I_R}\\
    I_{I} &= \frac{I}{V}\text{cos}\left(\phi\right)V_{I} - \frac{I}{V}\text{sin}\left(\phi\right)V_{R} = I_{GI} - I_{BI} \label{eq: I_I}
\end{align}

\begin{figure}[h!]
    \vspace{-0.175cm}
    \centering
    \begin{subfigure}[h]{0.135\textwidth}
        \centering  
        \includegraphics[width=\textwidth]{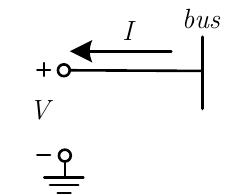}
        \caption{\hspace*{0.25em}}  
        \label{fig:RTU}
    \end{subfigure}
    \hfill
    \begin{subfigure}[h]{0.165\textwidth}  
        \centering 
        \includegraphics[width=\textwidth]{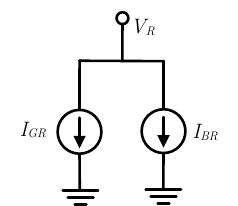}
        \caption{\hspace*{0.25em}}  
        \label{fig:RTU_Real}
    \end{subfigure}
    \hfill
        \begin{subfigure}[h]{0.165\textwidth}  
        \centering 
        \includegraphics[width=\textwidth]{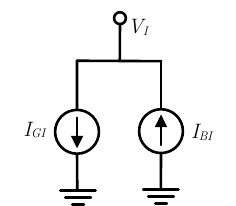}
        \caption{\hspace*{0.25em}}  
        \label{fig:RTU_Imag}
    \end{subfigure}
    \caption{RTU injection model: (a) measurement data; (b) real circuit; (c) imaginary circuit.}
    \label{fig:RTU_lin_model}
\vspace{-1em}
\end{figure}

\noindent Unlike the nonlinear circuit model from \cite{Jovicic} that represents RTU measurements as admittances, these dependencies are modelled as independent current sources as shown in Fig.~\ref{fig:RTU_lin_model}, but then include in the objective function a term that incentivizes the relationship with the voltage, as given in \eqref{eq: I_R} and \eqref{eq: I_I}. The model in Fig.~\ref{fig:RTU_lin_model} is given for the case where the reference directions of the measured voltage and current correspond to load conditions, i.e. consumption of active and reactive power. Therefore, active power generation renders $I_{GR}$ and $I_{GI}$ negative, while reactive power generation yields negative values for $I_{BR}$ and $I_{BI}$.

In order to estimate the most likely state of the system and accurately represent the measured data, the difference between each current source value $I_{RTU}$ (unknown variable) and the corresponding linear measurement function $h(x)$ needs to be minimized. Based on equations \eqref{eq: I_R} and \eqref{eq: I_I}, the corresponding current source-measurement function pairs are given in Table \ref{tab:RTU_pair}, where $V$, $I$ and $\phi$ are measured values.

\begin{table}[t!]
\vspace{0.2cm}
\centering
\caption{RTU Measurement Functions}
\label{tab:RTU_pair}
\begin{tabular}{ |c|c| }
 \hline
 $\boldsymbol{I_{RTU}}$ & $\boldsymbol{h(x)}$\\ 
 \hline
 $I_{GR}$ & $\frac{I}{V}\text{cos}\left(\phi\right)V_R$\\ 
 \hline
 $I_{BR}$ & $\frac{I}{V}\text{sin}\left(\phi\right)V_I$\\ 
 \hline
 $I_{GI}$ & $\frac{I}{V}\text{cos}\left(\phi\right)V_I$\\ 
 \hline
 $I_{BI}$ & $\frac{I}{V}\text{sin}\left(\phi\right)V_R$\\ 
 \hline
\end{tabular}
\vspace{-0.2cm}
\end{table}

As certain transformation of originally measured data is performed, the accompanying weight coefficients cannot be determined directly, as in the case of PMU measurements. The measurement functions are equal to the product of either real or imaginary bus voltages, which are unknown variables in the proposed formulation, and RTU measurements that are known quantities. Thus, for each measurement function $h(x)$, the corresponding weight coefficient is calculated as reciprocal value of the variance that is obtained based on the original measurements that appear in $h(x)$, their standard deviations and equation \eqref{eq: var_product}. In case line flow data are available instead of the injection current measurement, and they are measured in all lines incident to the respective bus, the injection current can be calculated by summing up all line currents. The variance of the so obtained injection current pseudo-measurement can be determined based on equation \eqref{eq: var_sum}.

Given that the circuit models for transmission lines, transformers and phase-shifters that are presented in \cite{Pandey} are linear, they will be used in this formulation in the same form. 
\subsection{Optimization Problem} \label{subsec: 3b}
Given the equivalent circuit of the system along with the measurements, an optimization problem is formulated in order to estimate the most probable state of the system based on measurement uncertainties and specific features of the PMU circuit model. For this purpose, the following problem is defined: 
\begin{subequations}
\vspace{-0.1cm}
\label{eq: optimization}
\begin{align}
\hspace{-2.4mm}\min &\sum_{j\in{\Omega_{PMU}}} \Big(I^2_{G_{PMU,R,j}} + I^2_{G_{PMU,I,j}} + \Delta V_{PMU,R,j}^2 \nonumber\\ &+ \Delta V_{PMU,I,j}^2 + \Delta I_{PMU,R,j}^2 + \Delta I_{PMU,I,j}^2 \Big) \nonumber\\&+\sum_{j\in{\Omega_{RTU}}} \Big(\Delta I_{GR,j}^2 + \Delta I_{BR,j}^2 + \Delta I_{GI,j}^2 + \Delta I_{BI,j}^2\Big) \label{eq: obj_func}
\end{align}
\vspace{-0.2cm}
\begin{equation}
    \text{s.t.} \quad\quad I_{ct}(\boldsymbol{X}) = 0 \label{eq: lin_eq}
\end{equation}
\end{subequations}
where $\Omega_{PMU}$ and $\Omega_{RTU}$ are the sets of PMU and RTU sub-circuits in the system, respectively. $\boldsymbol{X}$ is the vector of equivalent circuit state variables and is equal to:
\begin{equation}
    \boldsymbol{X} = [\boldsymbol{V_{BUS}},\,\boldsymbol{V_{PMU}},\,\boldsymbol{I_{PMU}},\,\boldsymbol{I_{RTU}},\,\boldsymbol{I_{G_{PMU}}},\,\boldsymbol{I_V}]^T
\end{equation}
where $\boldsymbol{V_{BUS}}$ is the vector of node voltages in rectangular form, $\boldsymbol{V_{PMU}}$ and $\boldsymbol{I_{PMU}}$ are vectors of PMU voltage and current variables as used for the independent sources, $\boldsymbol{I_{RTU}}$ represents the vector of RTU current variables as used in the respective independent current sources, $\boldsymbol{I_{G_{PMU}}}$ is the vector of currents flowing through the PMU conductances, and $\boldsymbol{I_{V}}$ comprises the currents through the PMU voltage sources.

The objective function given in \eqref{eq: obj_func} utilizes a weighted least square approach and comprises several terms of which each will be explained in the following. As it was previously discussed, the case of perfectly accurate measurements would cause an operating point of the equivalent circuit characterized by zero currents in the PMU conductances. The optimization problem \eqref{eq: optimization} minimizes the currents flowing through the PMU conductances ($I_{G_{PMU}}$), thus minimizing the measurement discrepancies and yielding conditions that correspond to the most likely operating point of the circuit, i.e. the power system. In general, the currents flowing through conductances $G_{PMU}$ can be expressed as:
\begin{equation}
    I_{G_{PMU},Z} = G_{PMU}(V_{PMU,Z}-V_{BUS,Z}), \,\,Z\in \{R,I\}
\end{equation}
where $V_{PMU}$ is the PMU voltage source connected to the first terminal of the respective conductance, while $V_{BUS}$ is the voltage at the second terminal, i.e. voltage at the bus where the corresponding PMU device is installed. Minimization of $I_{G_{PMU}}$ also reduces the difference between $V_{PMU}$ and $V_{BUS}$, thus enabling accurate modelling of the PMU measurements. Therefore, a high weight factor has to be assigned to the $I_{G_{PMU}}$ terms in \eqref{eq: obj_func}. Since a current flowing through a conductance is proportional to its value, $G_{PMU}$ can actually serve as a weight factor and its value is selected accordingly. 

Under the assumption that all measurements are normally distributed, a measured value also represents the most likely value, thus the difference between each variable in the PMU sub-circuit and its measured value should be minimized. General expressions for $\Delta V_{PMU,Z}$ and $\Delta I_{PMU,Z}$ with $Z\in\{R,I\}$ are:
\begin{align}
    \Delta V_{PMU,Z} &= \sqrt{W}\,\left(V_{PMU,Z} - \widetilde{V}_{PMU,Z}\right)\\
    \Delta I_{PMU,Z} &= \sqrt{W}\,\left(I_{PMU,Z} - \widetilde{I}_{PMU,Z}\right)
\end{align}
where $V_{PMU}$ and $I_{PMU}$ are PMU voltage and current variables, respectively, while $\widetilde{V}_{PMU}$ and $\widetilde{I}_{PMU}$ are their corresponding measured values. $W$ represents the weight coefficient calculated for the respective measurement. A similar approach is used for RTU sub-circuits. For each RTU current source the respective $\Delta I_{RTU}$ term is equal to:
\begin{equation}
    \Delta I_{RTU} = \sqrt{W}\left(I_{RTU} - h(x)\right)
\end{equation}
where $I_{RTU}$ represents the current variable for the source being observed, i.e. $I_{GR}$, $I_{BR}$, $I_{GI}$ or $I_{BI}$, while $h(x)$ is the measurement function correlated to the $I_{RTU}$, based on Table \ref{tab:RTU_pair}. $W$ is the weight factor calculated for the measurement function $h(x)$. 

The optimization problem given in \eqref{eq: optimization} is subject to a set of linear equality constraints \eqref{eq: lin_eq}, defined by the governing equations of the system's equivalent circuit. These can be formulated by implementing some of the frequently used circuit formulation methods, such as Modified Nodal Analysis. 

In order to solve for the optimal solution, the first-order necessary and sufficient optimality conditions have to be derived. The Lagrangian function for the optimization problem \eqref{eq: optimization} is defined as: 
\begin{equation}
    \mathcal{L}(\boldsymbol{X},\,\boldsymbol{\lambda}) = f(\boldsymbol{X}) + \boldsymbol{\lambda}^TI_{ct}(\boldsymbol{X}) 
\end{equation}
where $f(\boldsymbol{X})$ is the objective function \eqref{eq: obj_func} and $\boldsymbol{\lambda}$ is the vector of Lagrange multipliers. Then, the KKT conditions are obtained by differentiating $\mathcal{L}(\boldsymbol{X},\,\boldsymbol{\lambda})$ with respect to the primal ($\boldsymbol{X}$) and dual ($\boldsymbol{\lambda}$) variables. As a result, the following set of equations defines the KKT conditions for the problem at hand:
\begin{gather}
    \nabla_{\boldsymbol{X}}f(\boldsymbol{X}) + \nabla^T_{\boldsymbol{X}}I_{ct}(\boldsymbol{X})\boldsymbol{\lambda} = 0 \label{eq: set1}\\
    I_{ct}(\boldsymbol{X}) = 0 \label{eq: set2}
\end{gather}
where $\nabla_{\boldsymbol{X}}f(\boldsymbol{X})$ is the gradient vector of $f(\boldsymbol{X})$ and $\nabla_{\boldsymbol{X}}I_{ct}(\boldsymbol{X})$ is the Jacobian matrix of the set of equations. A very important observation is that the partial derivatives of $\mathcal{L}(\boldsymbol{X},\,\boldsymbol{\lambda})$ with respect to $\boldsymbol{X}$ are linear functions, as $\nabla_{\boldsymbol{X}}f(\boldsymbol{X})$ contains only linear terms and $\nabla_{\boldsymbol{X}}I_{ct}(\boldsymbol{X})$ is a constant matrix. As it was stated above, the set of circuit governing equations, contained in \eqref{eq: set2}, is also linear. Thus, the state of the system is obtained as a solution to the set of linear equations \eqref{eq: set1}-\eqref{eq: set2}. Consequently, the computational burden associated with the proposed formulation is very low, which lays the foundation for real-time state estimation even if standard PMU data reporting rates are considered, which can be as high as 100 frames per second \cite{Standard}.

\section{Results} \label{sec: 4}
The performance of the proposed linear formulation for power system state estimation is evaluated using the IEEE 14 and 118 bus test systems, the 2869 and 13659 bus systems provided by the PEGASE project \cite{13k_buses}, as well as the 70000 bus ARPA-E test case \cite{70k}.  

Both PMU and RTU measurements are used in each test case. The set of PMU measurements consists of voltage and current phasors in rectangular coordinates, while the set of RTU measurements comprises bus voltage magnitude, current magnitude and power factor data. It is assumed that each test system is observed predominantly by the set of conventional measurements, due to the small penetration of PMU devices in the present day power grids. The allocation of different types of measurement devices is presented in Table \ref{tab:measures}. In each system, a PMU is appointed to the slack bus in order to provide the reference for the angles. It is assumed that each PMU is capable of monitoring all lines incident to its bus. The same assumption is made for the RTU devices measuring line flows instead of injection currents. The chosen measurement allocation ensures full system observability for all test cases. 

\begin{table}[t!]
\centering
\caption{Measurement Allocation}
\label{tab:measures}
\begin{tabular}{ |c|c|c|c| }
 \hline
 \multicolumn{1}{|c|}{\multirow{2}{*}{\textbf{Test Case}}} & \multicolumn{1}{c|}{\multirow{2}{*}{\textbf{\# PMU buses}}} & \multicolumn{2}{c|}{\textbf{\# RTU Buses}}\\
 \cline{3-4}
  &  & \textbf{Injection} & \textbf{Flow}\\ 
 \hline
 14 buses & 3 & 6 & 5\\
 \hline
 118 buses & 10 & 58 & 50\\ 
 \hline
 2869 buses & 205 & 1176 & 1488\\ 
 \hline
 13659 buses & 779 & 6010 & 6870\\
 \hline
 70000 buses & 4135 & 30545 & 35320\\
 \hline
\end{tabular}
\vspace{-1.5em}
\end{table}

\begin{table}[b!]
\vspace{-0.4cm}
\centering
\caption{Measurement Standard Deviations}
\label{tab:deviations}
\begin{tabular}{|c|c|c|c|c|}
\hline
\multicolumn{3}{|c|}{\textbf{RTU}} & \multicolumn{2}{c|}{\textbf{PMU}}\\
\hline
Voltage   & Current   & Power Factor    & Voltage   & Current\\
 \hline
0.4\%   &  0.4\% & 0.5\%   & 0.02\%  & 0.02\%\\
\hline
\end{tabular}
\vspace{-1.5em}
\end{table}

All measurements of the same type are set to have the same standard deviation, and the selected values for measurement uncertainties are listed in Table \ref{tab:deviations}. Also, it is assumed that all measurements follow a Gaussian distribution with zero mean. The power flow solution of each system is considered to be its true operating point. Therefore, the measurement sets used in the test simulations are based on these values and the standard deviations from Table \ref{tab:deviations}. Each measurement is randomly selected within the range $[t-\sigma_t, t+\sigma_t]$, where $t$ is the true value, i.e. the accurate measurement, and $\sigma_t$ is its corresponding standard deviation.  

Test simulations are performed in Matlab. The states are estimated in rectangular form by solving a set of linear equations. The values of PMU conductances $G_{PMU}$ are set to 100 p.u., in order to ensure that a sufficiently large weight is assigned to the first term in the objective function \eqref{eq: obj_func}. This value is selected based on the difference between standard deviations of RTU and PMU measurements, and is set to be the same for all PMU conductances in the system, due to equal measurement uncertainties of all PMU measurements. 

\begin{figure}[t!]
\vspace{-0.2cm}
\centering
\begin{subfigure}{0.53\textwidth}
\includegraphics[width=\textwidth]{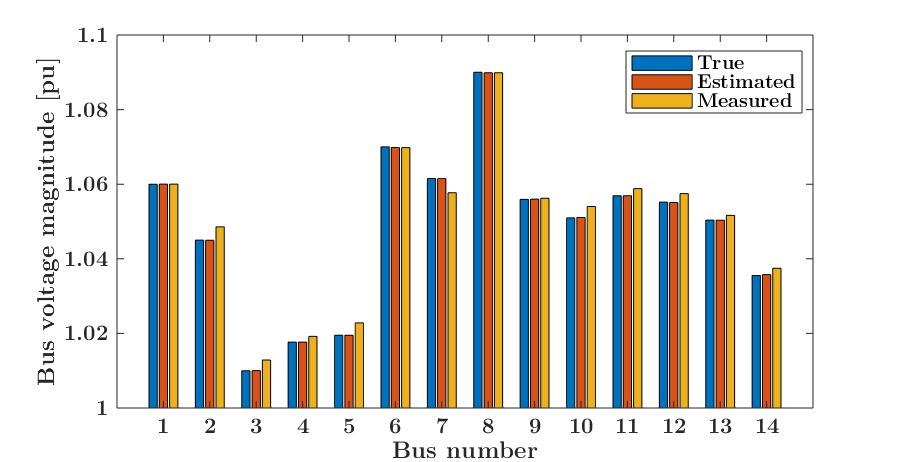}
\end{subfigure}
\caption{Comparison of true, estimated and measured values}
\label{fig:Comparison}
\end{figure}

In order to verify the capability of the proposed method to estimate the most probable states, it is applied to the 14 bus test case and the true, estimated and measured values of the bus voltage magnitudes are compared. In this particular test case, PMUs are assigned to buses 1, 6 and 8. Since all PMU voltage measurements and the voltage estimates are given in rectangular coordinates, they have to be translated to a polar form in order to enable the comparison. The results are presented in Fig.~\ref{fig:Comparison}. As can be seen, the proposed linear method provides estimates that are almost equal to the true values, and it definitely provides significantly better insight into the actual operating point of the system compared to the raw measurement data. 

To further prove the accuracy of the proposed formulation, the values of two performance indices are examined. The first index is the sum of squared differences between the true ($x^t$) and the estimated ($\hat{x}$) values of the system states:
\begin{equation}
    \sigma^2_{x} = \sum^{2N}_{i = 1}(\hat{x}_i - x^t_i)^2
\end{equation}
where $N$ is the number of buses. The second one represents the largest absolute discrepancy of an individual state estimate:
\begin{equation}
    \sigma_{max} = \max_{i\in{2N}}|\hat{x}_i - x^t_i|
\end{equation}
A total of 100 simulations are run for each test case in order to obtain results for various measurement values. In each simulation, random measurement values are sampled from the predefined range described earlier, while the allocation of different types of measurements is kept the same. 

The estimation results are presented in Table \ref{tab:results}. As can be seen, the proposed method provides very accurate state estimates, since both performance indices have very small values for all test cases. Particularly, the values of index $\sigma_{max}$ show that the estimation error is sufficiently small, even for the states that exhibit the largest discrepancy between the true and the estimated value. Furthermore, since the estimates are obtained by solving a set of linear equations, the computational time is extremely low, being significantly less than one second even for large test cases. Hence a solid foundation for real-time state estimation is developed. 
\begin{table}[t]
\centering
\caption{Average Performance Indices}
\label{tab:results}
\begin{tabular}{ |c|c|c| }
 \hline
 \textbf{Test Case} & $\boldsymbol{\sigma^2_x}$ & $\boldsymbol{\sigma_{max}}$\\ 
 \hline
 14 buses & 1.2803\,x\,$10^{-6}$ & 0.0006\\ 
 \hline
 118 buses & 9.8117\,x\,$10^{-5}$ & 0.0028\\ 
 \hline
 2869 buses & 1.2535\,x\,$10^{-3}$ & 0.0042\\ 
 \hline
 13659 buses & 0.0029 & 0.0058\\ 
 \hline
 70000 buses & 0.0593 & 0.0096\\ 
 \hline
\end{tabular}
\vspace{-0.4cm}
\end{table}

\section{Conclusion and Future Work} \label{sec: 5}
A novel linear formulation for the power system state estimation is introduced. It was shown that the ECF framework provides an alternative perspective on the state estimation problem, as both conventional and synchrophasor measurements were modelled by linear circuits. Both types of measurements were combined into a single-stage problem and states were estimated by solving a set of linear equations, thus rendering the computational time extremely low. The presented results obtained from a number of test cases demonstrate a high level of accuracy of the proposed method.

Circuit models for different combinations of measurement data will be derived in the future to extend the current modelling framework. Also, a possible application of the proposed linear formulation to common tools used in the power system monitoring area, such as bad data detection, will be explored. 



\begin{thebibliography}{00}
\bibitem{Abur}  A.  Abur  and  A.  G.  Exposito, \textit{Power System State Estimation: Theory and Implementation}. Marcel Dekker, Inc., 2004.
\bibitem{Schweppe} F. C. Schweppe and J.Wildes, ''Power system static-state estimation, part I: Exact model,'' \textit{IEEE Transactions on Power Apparatus and Systems}, vol. PAS-89, no. 1, pp. 120-125, Jan. 1970.
\bibitem{Phadke1985} A. G. Phadke, J. S. Thorp, and K. J. Karimi, ''Real time voltage phasor measurements for static state estimation,'' \textit{IEEE Transactions on PAS}, vol. 104, no. 11, pp. 3098-3107, Nov. 1985.
\bibitem{Ming2006} M. Zhou, V. A. Centeno, J. S. Thorp, and A. G. Phadke, ''An alternative for including phasor measurements in state estimators,'' \textit{IEEE Transactions on Power Systems}, vol. 21, no. 4, pp. 1930-1937, Nov. 2006.
\bibitem{Baltensperger2010} R. Baltensperger, A. Loosli, H. Sauvain, M. Zima, G. Andersson, and R. Nuqui, ''An implementation of two-stage hybrid state estimation with limited number of PMU,'' in \textit{10th IET International Conference on Developments in Power System Protection}, Mar. 2010.
\bibitem{Nuqui} R. F. Nuqui and A. G. Phadke, ''Hybrid linear state estimation utilizing synchronized phasor measurements,'' in \textit{IEEE PowerTech}, Jul. 2007.
\bibitem{Bi2008} T. S. Bi, X. H. Qin, and Q. X. Yang, ''A novel hybrid state estimator for including synchronized phasor measurements,'' \textit{Electric Power Systems Research}, vol. 78, no. 8, pp. 1343-1352, Aug. 2008.
\bibitem{Chakrabarti2010} S. Chakrabarti, E. Kyriakides, G. Ledwich, and A. Ghosh, ''Inclusion of {PMU} current phasor measurements in a power system state estimator,'' \textit{IET Generation, Transmission Distribution}, vol. 4, no. 10, pp. 1104-1115, Oct. 2010.
\bibitem{Valverde2011} G. Valverde, S. Chakrabarti, E. Kyriakides, and V. Terzija, ''A constrained formulation for hybrid state estimation,'' \textit{IEEE Transactions on Power Systems}, vol. 26, no. 3, pp. 1102-1109, Aug. 2011.
\bibitem{CMU2} M. Jereminov, D. M. Bromberg, X. Li, G. Hug, and L. Pileggi, ''Improving robustness and modeling generality for power flow analysis,'' in \textit{IEEE/PES Transmission and Distribution Conference and Exposition (T\&D)}, May 2016.
\bibitem{CMU3} M. Jereminov, D. M. Bromberg, A. Pandey, X. Li, G. Hug, and L. Pileggi, ''An equivalent circuit formulation for three-phase power flow analysis of distribution systems,'' in \textit{IEEE/PES Transmission and Distribution Conference and Exposition (T\&D)}, May 2016.
\bibitem{CMU4} A. Pandey, M. Jereminov, G. Hug, and L. Pileggi, ''Improving power flow robustness via circuit simulation methods,'' in \textit{IEEE PES General Meeting}, Jul. 2017.
\bibitem{Pandey} A. Pandey, ''Robust steady-state power grid analysis using equivalent circuit formulation with circuit simulation methods,'' Ph.D. dissertation, Carnegie Mellon University, 2018.
\bibitem{CMU6} A. Pandey, M. Jereminov, M. R. Wagner, G. Hug, and L. Pileggi, ''Robust convergence of power flow using Tx stepping method with equivalent circuit formulation,'' in \textit{2018 Power Systems Computation Conference (PSCC)}, Jun. 2018.
\bibitem{Jovicic} A. Jovicic, M. Jereminov, L. Pileggi, and G. Hug, ''An equivalent circuit formulation for power system state estimation including PMUs,'' in \textit{North American Power Symposium (NAPS)}, Sep. 2018.
\bibitem{Pileggi} L. Pileggi, R. Rohrer, and C. Visweswariah, \textit{Electronic Circuit \& System Simulation Methods.} McGraw-Hill, Inc., 1995.
\bibitem{Navidi} W. Navidi, \textit{Statistics for Engineers and Scientists.} New York : McGraw-Hill, 2015.
\bibitem{Caro} E. Caro and G. Valverde, ''Impact of transformer correlations in state estimation using the unscented transformation,'' \textit{IEEE Transactions on Power Systems}, vol. 29, no. 1, pp. 368-376, Jan. 2014.
\bibitem{standard} ''IEEE standard for synchrophasor measurements for power systems,'' \textit{IEEE Std C37.118.1-2011 (Revision of IEEE Std C37.118-2005)}, pp. 1-61, Dec. 2011.
\bibitem{13k_buses} C. Josz, S. Fliscounakis, J. Maeght, and P. Panciatici, ''AC Power Flow Data in MATPOWER and QCQP Format: iTesla, RTE Snapshots, and PEGASE,'' \textit{ArXiv e-prints}, Mar. 2016.
\bibitem{70k} A. B. Birchfield, T. Xu, and T. J. Overbye, ''Power flow convergence and reactive power planning in the creation of large synthetic grids,'' \textit{IEEE Transactions on Power Systems}, vol. 33, no. 6, pp. 6667-6674, Nov. 2018.
\end{thebibliography}
\end{document}